\documentclass{article}

\usepackage{PRIMEarxiv}
\usepackage{amsmath}
\usepackage[utf8]{inputenc} % allow utf-8 input
\usepackage[T1]{fontenc}    % use 8-bit T1 fonts
\usepackage{hyperref}       % hyperlinks
\usepackage{url}            % simple URL typesetting
\usepackage{booktabs}       % professional-quality tables
\usepackage{amsfonts}       % blackboard math symbols
\usepackage{nicefrac}       % compact symbols for 1/2, etc.
\usepackage{microtype}      % microtypography
\usepackage{lipsum}
\usepackage{fancyhdr}       % header
\usepackage{graphicx}       % graphics
\graphicspath{{media/}}     % organize your images and other figures under media/ folder

\title{Exploring rhythm formant analysis for Indic language classification
\thanks{\textit{\underline{Citation}}: 
\textbf{Authors. Title. Pages.... DOI:000000/11111.}} 
}

\author{
  Parismita Gogoi \\
  IIT Guwahati, India \\
  DUIET, Dibrugarh University, India \\
  \texttt{parismitagogoi@iitg.ac.in} \\
   \And
   Sishir Kalita \\
  Armsoftech.air, India \\
  \texttt{sisiitg@gmail.com} \\
 \And
   Priyankoo Sarmah \\
  IIT Guwahati, India \\
  \texttt{priyankoo@iitg.ac.in} \\
 \And
     S.R Mahadeva Prasanna \\
  IIT Dharwad, India \\
  IIIT Dharwad, India \\
  \texttt{prasanna@iitdh.ac.in} \\
}

\begin{document}
\maketitle

\begin{abstract}
This paper reports a preliminary study on quantitative frequency domain rhythm cues for classifying five Indian languages: Bengali, Kannada, Malayalam, Marathi, and Tamil. We employ rhythm formant (R-formants) analysis, a technique introduced by Gibbon that utilizes low-frequency spectral analysis of amplitude modulation and frequency modulation envelopes to characterize speech rhythm. Various measures are computed from the LF spectrum, including R-formants, discrete cosine transform-based measures, and spectral measures. Results show that threshold-based and spectral features outperform directly computed R-formants. Temporal pattern of rhythm derived from LF spectrograms provides better language-discriminating cues. Combining all derived features we achieve an accuracy of 69.21\% and a weighted F1 score of 69.18\% in classifying the five languages. This study demonstrates the potential of RFA in characterizing speech rhythm for Indian language classification.
\end{abstract}
\section{Introduction}
The analysis and understanding of speech rhythm are important for a variety of speech applications, including language/dialect identification. With the use of machine learning techniques and statistical models, speech rhythm analysis through signal processing has advanced significantly. More recent attention has been on the development of an inductive long-term rhythm analysis approach known as the low-frequency (LF) rhythm formant analysis (RFA), proposed by Gibbon~\cite{Gibbon2019quantify, gibbon2021jipa, gibbon18_speechprosody}. Quantitative frequency domain rhythm analysis supports the idea of LF spectral analysis of long-term amplitude modulation (AM) and frequency modulation (FM) envelopes. This frequency domain method doesn't need the linguistic units in the speech signal to be explicitly labeled like annotation-based deductive methods do, like, rhythm metrics~\cite{RAMUS,dauer1983stress,gibbonRhythmZoneTheory}. Moreover, these annotation-based approaches may be challenging for studying the rhythm for low-resource languages with less text responses, long duration speech files, spontaneous speech, etc. The beauty of RFA is that it allows for analysis of rhythm variation in larger utterances rather than restricting it to words, phrases, and sentences.

In RFA, Gibbon conceptualized the idea of rhythm formants (R-formants)~\cite{gibbon18_speechprosody,gibbonRhythmZoneTheory,gibbon2021jipa}, which are frequency values corresponding to the magnitude peaks in the LF spectrum. The amplitude envelope conveys LF information, establishing a rhythmic hierarchy that supports spoken language structure. Changes in physical speech rhythm dynamics are examined by correlating to various linguistic units, such as syllables and words~\cite{gibbon2021jipa}. In general, AM corresponds to the syllable sonority contour in the signal, whereas FM conveys several linguistic information, such as lexical tone, pitch accent, and intonation~\cite{gibbonComp, gibbon18_speechprosody}.

The LF spectrum calculated from the entire utterance can only provide gross rhythm information. Therefore, RFA further explores the possibilities of studying the rhythm variation over time using the LF spectrogram. The rhythm spectrogram derived from the AM and FM envelopes and the rhythm formant vector across the spectrogram are explored in story readings by ten Mandarin native speakers~\cite{gibbon2020storyreading}. Furthermore, temporal information about long-term rhythm is investigated using the contour of the frequencies of amplitude peaks across the LF rhythm spectrogram~\cite{gibbon2021jipa}. This information is further utilized to classify readings in English, Mandarin, and German.

The RFA findings indicate that analyzing the rhythm formant characteristics in the AM and FM spectra may lead to valuable research on the phonetics of prosodic typology. This research can be applied to analyze language, dialect, or speech style. Gibbon's exploratory work suggests the potential of utilizing RFA for language classification. However, there has been no attempt to use RFA to derive cues for Indian language classification. The current state-of-the-art, efficient language identification systems are based on deep learning models. The conventional language identification system uses SDC as features and language representation vectors, such as i-vector and x-vector with TDNN architectures~\cite{dey2022overview, martinez2012ivector, abdurrahman2021spoken}. Recently, researchers have started exploring the self-supervised based representations for classifying the languages~\cite{boito2024mhubert, bartley2023accidental, arora2024evaluation}.

We attempt to investigate the RFA by extracting rhythm correlates derived from the LF spectrum and LF spectrogram. This preliminary investigation aims to understand the usability of cues derived from the RFA for language discrimination. Our focus is not on developing a high-performing language identification model, but on analyzing the extent to which AM temporal envelope and FM envelope can provide language discriminative cues through their LF spectral analysis.

\section{Database}
Indic Speech Databases developed at IIIT Hyderabad, India, have been used in the present work~\cite{Prahallad2012TheII}. The IIIT-H Indic speech databases is composed of text and speech data of 1000 sentences each in Bengali, Hindi, Kannada, Malayalam, Marathi, Tamil, and Telugu respectively. The selected languages has over 10,000 Wikipedia entries as textual resources, and native speakers record the speech in their preferred dialect. For our experiment, we have considered the speech files, which have a duration of more than 4 sec. This is done because in the analysis we have used a 3-second moving window to compute the LF spectrogram (details are given in Section~\ref{spectrogramm}). After all the processing, we get five languages for which we have a comparable number of files. Therefore, we are using these five languages for our preliminary investigation. RFA for language classification are, viz., Bengali (ben), Kannada (kan), Malayalam (mal), Marathi (mar), and Tamil (tam).

\section{Methodology}
This section discusses the procedure for deriving LF spectrum and LF spectrograms from the AM and FM envelopes and computing the rhythm cues from them~\cite{gibbon2022sp,gibbonComp,Parismita}. \subsection{LF spectrum based measures}\label{rf1}
The RFA enables the analysis of the dynamics of natural spontaneous speech in terms of rhythm variation by applying the concept of regular oscillations below 10 Hz.
The method for deriving the LF spectrum and R-formants from the AM envelope involves several key steps~\cite{gibbon2021jipa,gibbonComp}. First, the speech signal is normalized, and its AM envelope is computed using the absolute Hilbert transform. This envelope is then smoothened, and its Fast Fourier transform is calculated to obtain the LF spectrum, and consider the spectrum in the frequency range 0-10 Hz. After removing the DC component and normalizing the spectrum amplitude, dominant magnitude peaks are detected using a peak peaking algorithm implemented on the SciPy~\cite{2020SciPy} Python package, resulting in R-formants characterized by their frequencies.

\begin{figure}[t]
\begin{center}
\includegraphics[scale = 0.8]{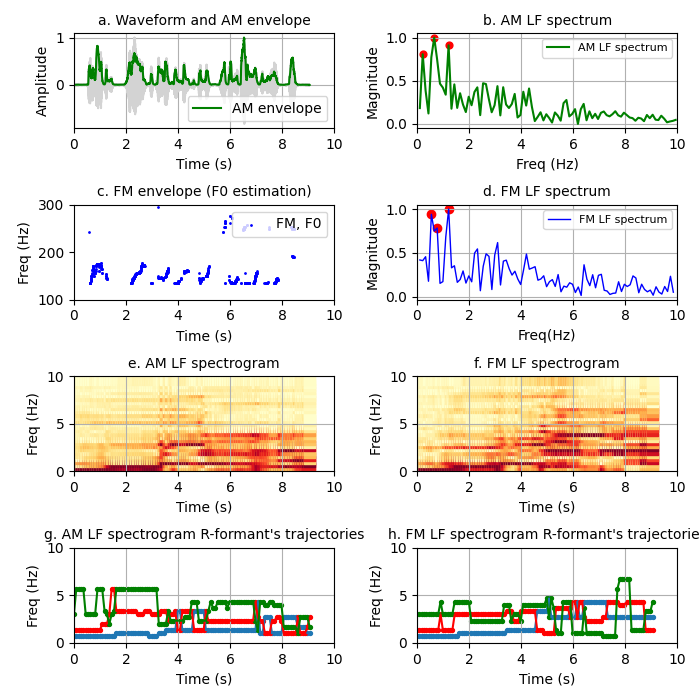}
\caption{Illustration of LF spectrum, LF spectrogram, and temporal trajectories for three dominant R-formants for AM and FM envelopes. a) Speech waveform and AM envelope, b) AM LF spectrum, c) F0 contour, d) FM LF spectrum, e) AM LF spectrogram, f) FM LF spectrogram, g) three dominant R-formant trajectories computed from AM LF spectrogram, and h) three dominant R-formant trajectories computed from AM LF spectrogram.}\label{fig:fmammain}
\end{center}
\end{figure}

For computing the LF spectrum and R-formants from the FM envelope, it also has similar steps as the AM envelope, except for the computation of the FM envelope in Step 2. The fundamental frequency (F0) contour of the speech signal is extracted followed by a smoothing of the F0 contour to obtain the FM envelope. We utilize the RAPT pitch tracking algorithm~\cite{talkin1995robust}. This algorithm's implementation is drawn from the $\textit{pysptk}$~\cite{yamamoto2019r9y9} Python package. 
To facilitate a more accurate spectral analysis, we adjust the FM envelope to align with the median F0~\cite{gibbon2021jipa}. However, while plotting the figure (Figure~\ref{fig:fmammain} (c)), discontinuities due to voiceless consonants and pauses are normalized to zero in the FM envelope.

Initially, we have investigated the R-formants computed from the LF spectrum for language classification. We explore the first six dominant R-formants for the analysis, as Gibbon mentioned in~\cite{gibbon18_speechprosody, gibbon2021jipa}. Figure~\ref{fig:fmammain} demonstrates how the R-formants are derived from audio clips of Malayalam speakers. We have adapted the codes from Gibbon's RFA Github repository~\footnote{https://github.com/dafyddg/RFA} to compute the AM and FM envelopes and R-formants. Figure~\ref{fig:fmammain} (a) depicts the speech waveform and its amplitude envelope. A positive amplitude envelope outlines the upper positive half of the waveform (green color). The top rightmost plot (Figure~\ref{fig:fmammain} (b)) displays the amplitude spectrum derived using FFT and R-formants (red dots) for the 3 most prominent frequencies derived by the method mentioned above for visualization. The R-formants computed from the FM envelope are shown in Figure~\ref{fig:fmammain} (d).
\subsubsection{Threshold-based measures}
As discussed in our previous work on Mising and Assamese rhythm discrimination~\cite{Parismita}, if the R-formants are derived from the LF spectrum-based threshold, it provides better performance than the most dominant R-formants. A detailed description of the threshold-based feature computation and its motivation can be found in~\cite{Parismita}. Next, we compute three features from the LF spectrum of AM and FM: (1) the number of detected peaks (NDP), (2) the mean of the detected peaks' frequency values (MFDP), and (3) the variance of the detected peaks' frequency values (VFDP). We expect that these features will represent the gross structure of the rhythm of a speech utterance.
\subsubsection{DCT-based characterization of LF spectrum}
The above-mentioned LF spectrum's threshold-based characterization may not capture the spectrum's shape~\cite{Parismita}. We hypothesize that the shape of the LF spectrum may have language-specific variations.
Hence, we explore coefficients derived from the discrete cosine transform (DCT) of the LF spectrum using Equation~\ref{eq_dct}. Moreover, DCT will provide a compact representation of the LF spectrum. We have considered the first four DCT coefficients in this exploratory work.
\begin{equation}\label{eq_dct}
X_k = \sum_{n=0}^{N-1}x_n cos[\frac{\pi}{N} (n + \frac{1}{2})k], \text{for \textit{k} = \textit{0} ... \textit{N}-\textit{1}}
\end{equation}
\subsubsection{Computation of spectral measures}
We have analyzed the frequency distribution patterns in the LF spectrum using the seven spectral measures, viz., spectral centroid, spread, rolloff, flatness, entropy, skewness, and kurtosis. A detailed description of the computation of these measures with equations can be found in~\cite{he2022characterizing}.
\subsection{LF spectrogram based measures}\label{spectrogramm}
As discussed earlier, Gibbon proposed the utilization of LF spectrograms for analyzing the long-term rhythm in speech. For the computation of LF spectrograms, the AM and FM envelopes need to be computed as discussed earlier. A moving window of duration 3 sec~\cite{gibbon2021jipa} is used to compute the fast Fourier transform of the AM or FM envelope. This allows us to see subtle and progressive variations in the rhythm~\cite{gibbon2021jipa}. The resulting LF spectrum, computed from each 3-sec chunk of AM/FM envelope, is stacked to form a time-frequency representation of long-term rhythm variation, also known as the AM/FM LF spectrograms (Figure~\ref{fig:fmammain} (e)). The LF spectrograms are computed within the 0 to 10 Hz frequency range. We discard the 0 Hz component (DC part) and normalize the spectrum's amplitude. Spectral frequencies having the six highest magnitude peak values are extracted for each LF spectrum, which is termed rhythm formant (R-formant)~\cite{gibbon2020storyreading,gibbonComp}, and the R-formant trajectories are composed along the time axis. Also, the trajectories of six magnitude values are also computed. We study the variance of the R-formant and magnitude trajectory~\cite{gibbon2022sp} to observe rhythmic variations between the five Indian languages for both AM and FM envelopes. Figure~\ref{fig:fmammain} (g) and (h) show the three dominant R-formant trajectories for visualization computed from AM and FM LF spectrograms, respectively. For our experiments, we have considered the first six rhythm formants. We calculate the variance of each R-formant trajectory, resulting in 12 (6 from R-formants and 6 from magnitude values) variance-based rhythm values for each utterance.
\begin{table}[h!]
\centering
\caption{Performance of SVM-based classification using all the explored measures.}
\label{table:svm1}
{\resizebox{\textwidth}{!}{%
\begin{tabular}{|c|c|c|c|c|c|}
\hline
Envelope    & Features                              & Spectrum/Spectrogram   & Dimension         & Accuracy (\%)     & Weighted F1 score (\%)   \\ \hline
AM          & R-formants                            &
Spectrum                & 6                 & 36.49 & 38.30 \\ \hline
FM          & R-formants                            &
Spectrum                & 6                 & 33.23 & 35.0 \\ \hline
AM + FM          & R-formants                            &
Spectrum                & 6 + 6 = 12                 & 40.11 & 39.45 \\ \hline
AM          & LFDP+MFDP+VFDP+DCT features           & Spectrum               & 7                 & 47.19          & 46.22          \\ \hline
FM          & LFDP+MFDP+VFDP+DCT features           & Spectrum               & 7                 & 42.22          & 41.22          \\ \hline
AM + FM (A) & LFDP+MFDP+VFDP+DCT features           & Spectrum               & 7 + 7 = 14                & 53.31          & 53.35          \\ \hline
AM          & Spectral features                     & Spectrum               & 7                 & 48.2           & 49.27          \\ \hline
FM          & Spectral features                     & Spectrum               & 7                 & 45.02          & 46.39          \\ \hline
AM + FM (B) & Spectral features                     & Spectrum               & 7 + 7 = 14                & 54.66          & 54.86          \\ \hline
AM          & Variance of R-formants (VarRFs)       & Spectrogram            & 6                 & 44.41          & 45.27          \\ \hline
AM          & Variance of magnitude values (VarMag) & Spectrogram            & 6                 & 36.37          & 37.77          \\ \hline
AM          & VarRFs and VarMag                     & Spectrogram            & 6 + 6 = 12                & 50.22          & 51.12          \\ \hline
FM          & VarRFs                                & Spectrogram            & 6                 & 39.12          & 42.05          \\ \hline
FM          & VarMag                                & Spectrogram            & 6                 & 34.33          & 34.89          \\ \hline
FM          & VarRFs and VarMag                     & Spectrogram            & 6 + 6 = 12                & 42.79          & 43.72          \\ \hline
AM + FM (C) & VarRFs and VarMag                     & Spectrogram            & 24                & 58.12          & 58.43          \\ \hline
AM + FM     & A + B + C                             & Spectrum + Spectrogram & 24 + 14 + 14 = 52 & \textbf{69.21} & \textbf{69.18} \\ \hline
\end{tabular}}}
\end{table}

\subsection{SVM-based classification system}
We develop SVM-based classifiers to investigate the usefulness of LF-spectrum-based and LF-spectrum-based measures in language classification. In the SVM, derived features are fed as input for training and testing on the unseen test speech files from one of the languages~\cite{cortes1995support}. We split our dataset into two sets, namely the train set and the test set, where 80\% of the data is used for training and remaining for testing. We use the grid-search method to tune the hyperparameters of SVM based on cross-validation on the training set. The models are evaluated using accuracy and weighted F1-score evaluation metrics.
\section{Experimental results}
\begin{figure}[h!]
\begin{center}
\includegraphics[scale=1.3]{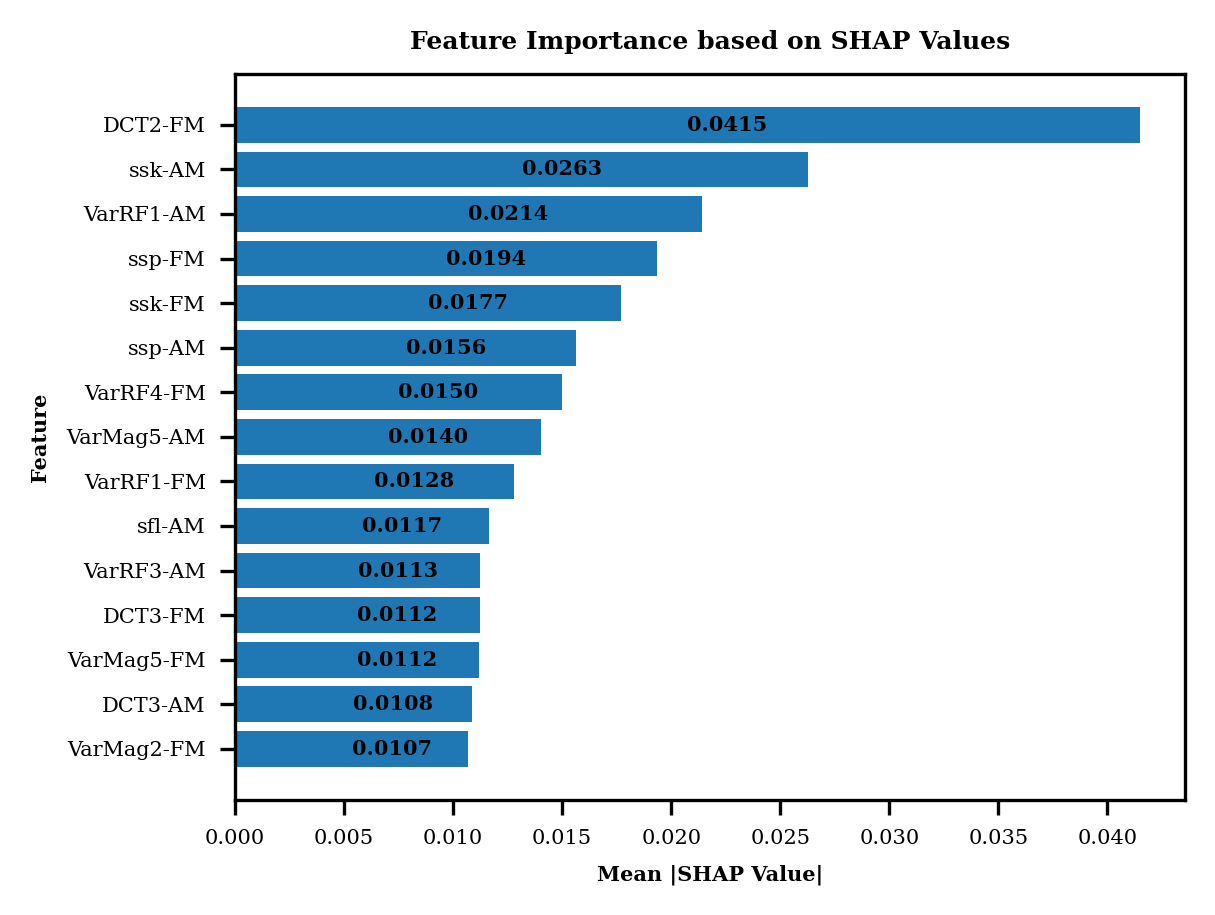}
\caption{Feature importance based on SHAP Values.}\label{fig:featimp}
\end{center}
\end{figure}
\begin{figure}[t]
\begin{center}
\includegraphics[scale=0.9]{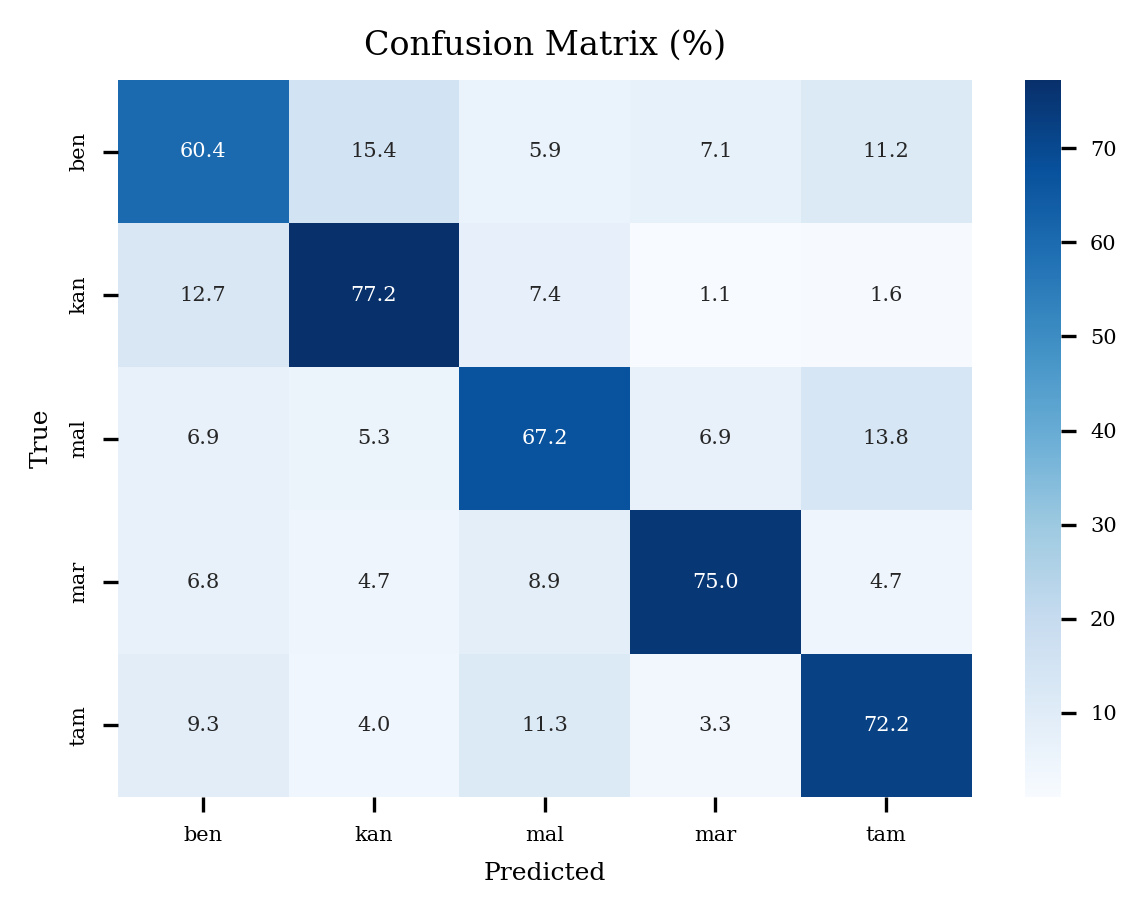}
\caption{Confusion matrix for the SVM-based five-class language classification using the 52-dimensional feature vector (A + B + C in Table~\ref{table:svm1}).}\label{fig:confmat}
\end{center}
%\vspace{-0.8cm}
\end{figure}
The results of the SVM-based classification are noted in Table~\ref{table:svm1}. According to the table, the accuracy and weighted F1 are 36.49\% and 38.30\%, respectively, with only R-formants. Furthermore, the performance of FM-based R-formants is lower than that of AM-based. While concatenating both AM and FM envelope-based R-formants, accuracy improved to 40.11\%. However, directly derived R-formants do not perform well in classifying the languages. It is evident from the table that threshold-based features provide performance improvements in comparison to directly computed R-formants for both AM and FM envelope-based LF spectrums. While combining the threshold-based features and DCT features for both AM and FM, performance increases to 53.31\% (around 12.31\% absolute performance improvement than R-formants). The table indicates that the spectral features, which model the distribution of the LF spectrum, outperform the threshold-based and DCT features. When combining the AM and FM-based spectral features, accuracy and weighted F1 score increased to 54.86\% and 54.66\%, respectively, and this result is best among all the features computed from the LF spectrum. The results reveal that both AM and FM possess complementary language discrimination information, and feature-level fusion enhances the performance.

The results of measures derived from the LF spectrogram are also mentioned in Table~\ref{table:svm1}. The variance of the R-formant's temporal trajectories consistently yields superior results compared to the variance of magnitude values for both the AM LF spectrogram and FM LF spectrogram. When VarRFs and VarMag are combined, we observe an accuracy of 50.22\% for AM, and it is the best result for measures derived for AM envelope. Moreover, when combined with FM-based measures, accuracy improves to 58.11\%. Therefore, the LF spectrogram's temporal variations of rhythm embed more language-discriminating cues than the gross LF spectrum for the entire utterance. Finally, when all the derived features, excluding the R-formants, are combined, we achieve an accuracy of 69.21\% and a weighted F1 score of 69.18\%, and the combined feature vector has a total dimension of 52. The feature importance based on SHAP (SHapley Additive exPlanations) values is shown in Figure~\ref{fig:featimp} for the top 15 important features. The figure reveals that while FM envelope-based measures did not perform well collectively in the performance evaluation, they accounted for 8 out of 15 top contributors at the individual level. Four of those eight measures, derived from the FM-LF spectrogram, demonstrate the significance of temporal rhythm variations in language discrimination. The DCT2-FM shows the highest individual contribution in classifying the five Indian languages. Spectral measures such as spectral skewness, spectral spread, and spectral flux are among the most important features based on the SHAP analysis. DCT measures, which represent the mean, curvature, and other higher-order fluctuations, model the language-discriminating cues from the LF spectrum. The confusion matrix for the five-class language classification for the 52-dimensional feature vector (A + B + C in Table~\ref{table:svm1}) is shown in~\ref{fig:confmat}. From the confusion matrix, it can be seen that Tamil is mostly confused with Malayalam and vice- versa, and this is maybe due to the fact that both are from the same language family. The 8.9\% of Marathi test data is confused with Malayalam. However, Kannada is confused with Bengali, and a proper justification for this behavior needs to be analyzed in future research.
\section{Summary and future directions}
This study explored Rhythm Formant Analysis (RFA) for classifying five Indian languages using low-frequency spectral analysis of AM and FM envelopes. We extracted features from LF spectra and spectrograms, computed various measures including R-formants and spectral features, and developed an SVM-based classifier. Our findings showed that threshold-based and spectral features outperformed direct R-formants, while temporal variations from LF spectrograms provided strong language-discriminating cues. Combining all derived features (excluding R-formants) achieved the highest accuracy of 69.21\%. Future work could expand this approach to more languages and larger datasets to validate its scalability. Additionally, applying deep learning techniques to learn rhythm cues directly from LF spectra and LF spectrograms could potentially improve classification performance. Further research could also investigate how these RFA-based features align with traditional rhythm typology classifications and explore their integration into broader speech technology applications.

\bibliographystyle{unsrt}  
\bibliography{mybib}

\end{document}